\documentclass{ws-ijmpcs}

\usepackage{relsize}
\def\trimmarks{}

\begin{document}

\markboth{A.~Courtoy $\&$ S. ~Liuti}
{The strong couplong constant from quark-hadron duality.}

%
\catchline{}{}{}{}{}
%

\title{ANALYSIS OF $\alpha_s$ FROM THE REALIZATION OF QUARK-HADRON DUALITY}

\author{A.~COURTOY\footnote{
Speaker.}}

\address{IFPA, AGO Department, Universite de Liege\\
Liege, Belgium.\\
INFN-LNF, Frascati, Italy.\\
aurore.courtoy@ulg.ac.be}

\author{SIMONETTA LIUTI}

\address{{\it Department of Physics, University of Virginia, 382 McCormick Rd. \\
Charlottesville, VA 22904, USA
\\
sl4y@virginia.edu}}

\maketitle

\begin{history}
\received{Day Month Year}
\revised{Day Month Year}
\end{history}

\begin{abstract}
We present an analysis of the role of the running coupling constant at the intersection of perturbative and nonperturbative QCD in the context of the quark-hadron duality \`a la Bloom-Gilman. 
Our framework will be  the unpolarized structure function of the proton  in the resonance region. We suggest that the realization of duality is related to the inclusion of nonperturbative effects at the level of the coupling constant.
The outcome of our  analysis is a smooth transition from perturbative to nonperturbative QCD physics, embodied in the running of the coupling constant at intermediate scales.
\keywords{Keyword1; keyword2; keyword3.}
\end{abstract}

\ccode{PACS numbers: 11.25.Hf, 123.1K}

\section{ Introduction}

QCD, the theory of strong interactions, which color singlets are the hadrons, is formulated in terms of quarks and gluons. The underlying paradox is that, so far, the physicists have failed to describe hadrons within QCD, due to the property of confinement. On the other hand,  hadron phenomenology is rich, while our current understanding prevents us from unveiling the hadron structure. Actually, at moderate energy scales, QCD tells us that  asymptotic freedom and confinement properties meet. Then  the partonic description develops into the  hadronic representation. The realization of this development from a perturbative to a nonperturbative description is still obscure. Various models of parton dynamics at low energies, mimicking QCD, have been proposed, and  they have led the way toward improvements in the understanding of hadron phenomenology. The other way round, purely phenomenological and perturbative approaches have been pursued as well.
For example, we know that Deep Inelastic processes allow us to look with a good resolution inside the hadron and  to resolve the very short distances, {\it i.e.} small configurations of quarks and gluons. 
At such short distances, one singles out a hard scattering process described through Perturbative QCD (pQCD). The large distance part of the process, {\it i.e.} the Parton Distribution Functions (PDFs), reflects how the quarks and gluons are distributed inside the target. PDFs have been evaluated within models, which validity is restricted to moderate energies, as well as parametrized from higher energy data, providing a complementary description.

Although, as we have just explained, the perturbative stage of a hard collision is distinct from the nonperturbative regime characterizing the hadron structure, early experimental observations suggest that, in specific kinematical regimes, both the perturbative and nonperturbative stages arise almost ubiquitously, in the sense that the nonperturbative description follows the perturbative one. 
There exists, in Deep Inelastic processes, a dual description between low-energy and high-energy behavior of a same observable, {\it i.e.} the unpolarized structure functions. Bloom and Gilman observed a connection between the structure function $\nu W_2(\nu, Q^2)$ in the nucleon resonance region and that in the deep inelastic continuum~[\refcite{Bloom:1970xb,Bloom:1971ye}]. The resonance structure function was found to be equivalent  to the deep inelastic one, when averaged over the same range in the scaling variable.  This concept is known as {\it parton-hadron duality}: the resonances are not a separate entity but are an intrinsic part of the scaling behavior of $\nu W_2$.
The meaning of duality is more intriguing when the equality between resonances and scaling happens at a same scale. 
It can be understood as a natural continuation of  the perturbative to the nonperturbative representation.

In this contribution to the proceedings, we study the Bloom-Gilman duality from a purely pertubative point of view, by analysing the scaling behavior of the resonances at the same low-$Q^2$, high-$x$ values as the $F_2$ data from JLab. We start from the analysis of Ref.~[\refcite{Liuti:2011rw}], where the implications of  parton-hadron duality are explored in the large $x$ region of inclusive electron proton scattering experiments.

Our study leads to an analysis of the role of the running coupling constant in the infrared region in tuning the experimental data~[\refcite{Liuti:2011rw,Bianchi:2003hi}]. 
The new approach on the freezing of the running coupling constant is outlined in Ref.~[\refcite{Courtoy:2013qca}].

\section{Quark-Hadron Duality in Electron-Proton Scattering}
\label{sec:duality}

Bloom--Gilman duality implies a one-to-one correspondence between the behavior of the structure function, $F_2$, for unpolarized electron proton scattering in the resonance region, and in the pQCD regulated scaling region. 
In Deep Inelastic Scattering (DIS), the relevant kinematical variables are the Bjorken scaling variable, $x=Q^2/2M\nu$ with $M$ being the proton mass and $\nu$ 
the energy transfer in the lab system, the four-momentum transfer, $Q^2$, and the invariant mass for the proton, $P$, for the virtual photon, $q$, and for the system, 
%
$W^2=(P+q)^2= Q^2\left(1-x\right)/x+M^2$.
%

For
large values of Bjorken $x  \geq 0.5$, and $Q^2$ in the multi-GeV$^2$ region, the cross section is dominated by resonance formation, {\it i.e.} $W^2 \leq 5$ GeV$^2$.
While it is impossible to reconstruct the detailed structure of the proton resonances, these remarkably follow the pQCD predictions when averaged over  the resonance region.

Bloom--Gilman duality was observed at the inception of QCD, its theoretical formulation through the OPE followed the emergence of QCD as the theory of the strong interactions. In the framework of structure functions, the OPE is an expansion in  twist.  The OPE formulation of quark-hadron duality~[\refcite{Rujula:1976tz}] suggests that the higher-twist contributions to the scaling structure function would either be small or cancel  otherwise duality would be strongly violated. 
However, the role of the higher-twist terms is still unclear since they would otherwise be expected to 
dominate the cross section at $x \rightarrow 1$.

To answer the question of the nature of a dual description, two complementary approaches have been adopted. The first is the nonperturbative model's view on the scaling of the structure functions at low-energies ; the second approach consists in a purely perturbative analysis from pQCD evolution. Although Bloom-Gilman duality has been known for years, quantitative analyses could be attempted only more recently, having at disposal the extensive, high precision data from Jefferson Lab~[\refcite{Melnitchouk:2005zr,Liang:2004tj}]. Perturbative QCD-based studies~[\refcite{Liuti:2011rw,Bianchi:2003hi,Liuti:2001qk}], have been presented that include  higher-twist contributions or, more generally, the evidence for nonperturbative inserts, which are required to achieve a fully quantitative fit, especially at large-$x$. It is the second approach that we will follow here.

A quantitative definition of  duality is accomplished by comparing limited intervals (integrated in Bjorken-$x$ over the entire resonance region, namely \emph {global duality}) defined according to the experimental data. 
Hence, we analyse the scaling results as a theoretical counterpart, or an output of pQCD, in the same kinematical intervals and at the same scale $Q^2$ as the data for $F_2$. It is easily realized that the ratio,
\begin{eqnarray}
R^{\mbox{\tiny exp/th}}(Q^2)&=&\frac{
\int_{x_{\mbox {\tiny min}}}^{x_{\mbox {\tiny max}}} dx\,
F_2^{\mbox {\tiny exp}} (x, Q^2)
}
{\int_{x_{\mbox{\tiny min}}}^{x_{\mbox{\tiny max}}} dx\,
F_2^{\mbox {\tiny th}} (x, Q^2)
}\quad,
\label{eq:ratio_1}
\end{eqnarray}
is equal to $1$ if duality is fulfilled. In the present analysis, we use, for $F_2^{\mbox {\tiny exp}} $, the data from JLab (Hall C, E94110)~[\refcite{Liang:2004tj}] reanalyzed (binning in $Q^2$ and $x$) as explained in~[\refcite{Monaghan:2012et}] as well as the SLAC data~[\refcite{Whitlow:1991uw}]. 

In our analysis, we use  different order expansions:
\begin{itemize}
\item the expansion in $\alpha_s$ is  performed here to  \emph{next-to-leading-order} in $\alpha_s$,
\item the expansion in twist includes here  \emph{leading-twist} PDFs,
\item the expansion in logarithms is considered, here, to \emph{all logarithms} for the expansion of $\alpha_{s}$.
\end{itemize}

The definition of the scaling structure function in Eq.~(\ref{eq:ratio_1}) relies on the fact that the pQCD evaluation is very well constrained in the region of interest ($x \gtrsim 0.2$) despite it does not correspond directly to measured data. $F_2^{\mbox {\tiny th}}$ is an input that once fed into the evolution equations determines the structure functions behavior at much larger $Q^2$,
\begin{eqnarray}
F_2^{NS} (x, Q^2) 
&=&x q(x,Q^2)+ \frac{\alpha_s}{4\pi} \mathlarger{\sum}_q \mathlarger{\int}_x^1 dz \, B_{\mbox{\tiny NS}}^q(z) \, \frac{x}{z}\, q\left(\frac{x}{z},Q^2\right)\quad,
\label{eq:convo}
\end{eqnarray}
where we have considered only the non-singlet (NS) contribution to $F_2$ since 
only valence quarks distributions are relevant in our kinematics. The PDFs, $q(x, Q^2)$, are  evolved to NLO to $Q^2$ from the initial scale $Q_0^2=1$GeV$^2$. We have chosen the MSTW08 set to NLO as initial parametrization~[\refcite{Martin:2009iq}].
The function $B_{\mbox{\tiny NS}}^q$ is the Wilson coefficient function for quark-quark.
As shown on Fig.~\ref{fig:ratio_lxr}, the ratio $R$ is not $1$ when considering only pQCD.

At finite $Q^2$, the effects of the target and quark masses modify the identification of the Bjorken variable with the light-cone momentum fraction. For massless quarks, the parton light-cone fraction is given by the Nachtmann variable $\xi$,
\begin{eqnarray}
\xi=\frac{2x}{1+r}\qquad \mbox{with}\qquad r=\sqrt{1+\frac{4x^2M^2}{Q^2}}\quad.
\end{eqnarray}
Theses corrections,
due to the finite mass of the initial nucleon, called TMCs, are included directly in $F_2^{NS}$ as  [\refcite{Georgi:1976ve}],%
\begin{eqnarray}
\label{TMC}
F_{2}^{NS (TMC)}(x,Q^2) & = &
    \frac{x^2}{\xi^2\gamma^3}F_2^{\mathrm{\infty}}(\xi,Q^2) + 
    6\frac{x^3M^2}{Q^2\gamma^4}\int_\xi^1\frac{d \xi'}{{\xi'}^2} 
F_2^{\mathrm{\infty}}(\xi',Q^2),
\end{eqnarray}
where $F_2^{\infty} \equiv F_2^{NS}$
is the structure function in the absence of TMC. 

TMCs move the ratio closer to unity, as represented by the open green diamonds in Fig.~\ref{fig:ratio_lxr}. 
At this stage, by including only TMCs and standard PDF parametrizations, we still observe  a large discrepancy with the data.

One possible explanation for the apparent violation of duality is the lack of accuracy in the PDF parametrizations at large-$x$. Therefore, the behavior of the nucleon structure functions in the 
resonance region needs to be addressed in detail  
in order to be able to discuss 
theoretical predictions in the limit $x \rightarrow 1$. 
Analyses~[\refcite{Liuti:2001qk}]  show  that one is now able to unravel different sources of scaling violations affecting the structure functions, 
namely TMC (that we have discussed above), Large $x$ Resummation effects (LxR), and dynamical higher-twists, 
in addition to the standard NLO perturbative
evolution. As a result, contrarily to what originally deduced in 
{\it e.g.} Ref.~[\refcite{Niculescu:2000tk}], a more pronounced role of the higher-twist terms 
is obtained, pointing at the fact that duality, defined on the basis of a
dominance of single parton scattering, {\it i.e.} suppression of final state interactions, could indeed be broken. 

We here propose yet another interpretation of the apparent violation of duality, that does not invoke final state interactions directly.  NLO pQCD evolution at large $x$ is sensitive to Large $x$ Resummation (LxR) effects. The consequence of LxR is a shift of the scale at which $\alpha_s$ is calculated
to lower values, with increasing $z$. 
This introduces a  model dependence within the pQCD approach in that  the value of the QCD running coupling  in the 
infrared region is regulated by LxR so to satisfy duality.
In other words, LxR contains an additional freedom, gathered in the definition of the coupling constant, to tune the scaling structure functions, simultaneously suppressing the higher-twist effects. The higher-twist effects get, in fact, absorbed in the coupling's infrared behavior.

\section{Large-$x$ Resummation}
\label{sec:lxr}

Quark-hadron duality can be understood as follows:  the knowledge of perturbative QCD can be used to calculate nonperturbative QCD physics'observables~[\refcite{Poggio:1975af}]. 
However, when considering pQCD observables at low scales, we implicitly face an interpretation problem. Higher terms in the perturbative expansion of that observable need be taken into account, by definition. Rephrazing, it gives: we are trying to make up for the perturbative to nonperturbative QCD physics transition in our perturbative analysis. In the present approach, this phase transition is  fully included in the interpretation of the role of the running coupling constant, at the scale of transition instead.

Let us  further develop this idea.
LxR  arise formally from terms containing powers of 
$\ln (1-z)$, $z$ being the longitudinal 
variable in the evolution equations, that are present in 
the Wilson coefficient functions $B_{\mbox{\tiny NS}}^q(z)$, in Eq.~(\ref{eq:convo}).
 To NLO and in the $\overline{\mbox{MS}}$ scheme, the Wilson coefficient function for quarks reads,
\begin{eqnarray}
B_{\mbox{\tiny NS}}^q(z)&=& \left[
\hat{P}_{qq}^{(0)}(z)\, \left\{
\ln\left(\frac{1-z}{z}\right)-\frac{3}{2}
\right\}
+\mbox{E.P.}
\right]_+\quad,
\end{eqnarray}
where E.P. means end points and $[\ldots]_+$ denotes the standard plus-prescription. The function $\hat{P}_{qq}^{(0)}(z)$ is the LO splitting function for quark-quark.
The logarithmic terms, {\it i.e.}, $\ln(1-z)$,  in $B_{\mbox{\tiny NS}}^q(z)$ become very large at large $x$ values. They need to be 
resummed to all orders in $\alpha_s$. 
Resummation was first introduced by  
linking this issue to the definition of the correct kinematical variable that determines the 
phase space for  real gluon emission
at large $x$. This was found to be $\widetilde{W}^2 = Q^2(1-z)/z$, 
instead of $Q^2$~[\refcite{Amati:1980ch}].
As a result, the argument of the strong coupling constant becomes $z$-dependent~[\refcite{Roberts:1999gb}],
%
$\alpha_s(Q^2) \rightarrow \alpha_s\left(Q^2 \frac{(1-z)}{z}\right)$.
%

In this procedure, however, an ambiguity is introduced, related to the need of continuing 
the value of $\alpha_s$  
for low values of its argument, {\it i.e.} for $z \rightarrow 1$. 
Since the size of this ambiguity is of the same order as the higher-twist corrections, it has been considered, in a previous work~[\refcite{Niculescu:1999mw}], as a source of theoretical error or higher order effects. We propose  an enlightening analysis from which one can draw $\alpha_s$ for values of the scale in the infrared region. 
To do so, we investigate the effect induced by changing the argument of $\alpha_s$ on the behavior of the $\ln(1-z)$-terms in the convolution Eq.~(\ref{eq:convo}). We resum those terms as
\begin{eqnarray}
\ln(1-z)&=&\frac{1}{\alpha_{s, \mbox{\tiny LO}}(Q^2) }\int^{Q^2} d\ln Q^2\, \left[\alpha_{s, \mbox{\tiny LO}}(Q^2 (1-z)) -\alpha_{s, \mbox{\tiny LO}}(Q^2) \right]
\equiv \ln_{\mbox{\tiny LxR}}\quad,
\end{eqnarray}
 including the complete $z$ dependence of $\alpha_{s, \mbox{\tiny LO}}(\tilde W^2)$ to all logarithms.\footnote{The terms proportional to $\ln z$ are not divergent at $z\to 1$.}

To all logarithms, the convolution becomes
\begin{equation}
F_2^{NS,  \mbox{\tiny Resum}} (x, Q^2)  =x q(x,Q^2)+ \frac{\alpha_s}{4\pi} \mathlarger{\sum}_q \mathlarger{\int}_x^1 dz \, B_{\mbox{\tiny NS}}^{ \mbox{\tiny Resum}}(z) \, \frac{x}{z}\, q\left(\frac{x}{z},Q^2\right),
\label{eq:convo_eff}
\end{equation}   
where,
\begin{equation}
B_{\mbox{\tiny NS}}^{ \mbox{\tiny Resum}} =  B_{\mbox{\tiny NS}}^q(z) - \hat{P}_{qq}^{(0)}(z)\, \ln(1-z)+\hat{P}_{qq}^{(0)}(z)\,\ln_{\mbox{\tiny LxR}}.
\end{equation}

Using $F_2^{(1),\mbox {\tiny DIS}} $ in Eq.~(\ref{eq:ratio_1}), the ratio $R$ decreases substantially, even reaching values lower than 1. It is a consequence of the change of the argument of the running coupling constant. At fixed $Q^2$, under integration over $x<z<1$, the scale $Q^2\times(1-z)/z$ is shifted  and can reach low values, where the running of the coupling constant starts blowing up. At that stage, our analysis requires nonperturbative information.

\begin{figure}[h]
\includegraphics[scale= .75]{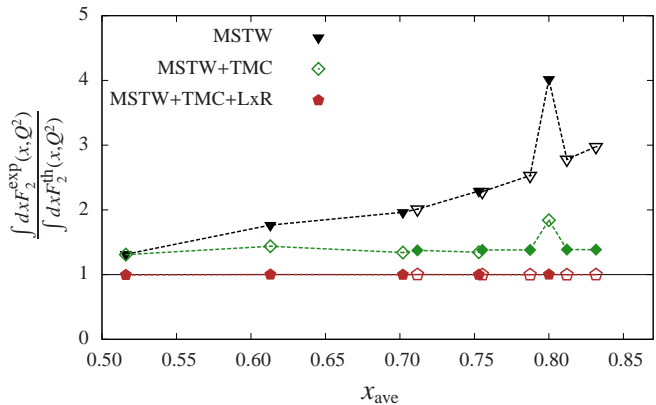}
\caption{The ratio $R^{\mbox{\tiny exp/th}}(x_{\mbox{\tiny ave}}, Q^2)$ at fixed $Q^2$, where the theoretical analysis have been performed with QCD evolution for the MSTW08 PDF set (black triangle) and MSTW08 PDF set plus target mass corrections (green diamonds). The red hexagons represent Eq.~(\ref{eq:ratio_max}). The legend is for JLab data, the opposite (open triangles, solid diamonds and open hexagons) correspond to SLAC data points.}
\label{fig:ratio_lxr}
\end{figure}

In the light of quark-hadron duality, it is necessary to prevent the evolution from enhancing the scaling contribution over the resonances. One possible way-out is to set a maximum value for the longitudinal momentum fraction, $z_{max}$, which defines a limit from which nonperturbative effects have to be accounted for. The functional form $\ln_{\mbox{\tiny LxR}}$ is therefore slightly changed. Two distinct regions can be studied: the ``running" behavior  in $x<z<z_{max}$ and the ``steady" behavior $z_{max}<z<1$. 
\begin{eqnarray}
F_2^{NS,  \mbox{\tiny Resum}} (x, z_{\mbox{\tiny max}}, Q^2)  &=&x q(x,Q^2)+ \frac{\alpha_s}{4\pi} \mathlarger{\sum}_q 
\left \{
\mathlarger{\int}_x^1 dz \, \left[B_{\mbox{\tiny NS}}^q(z) - \hat{P}_{qq}^{(0)}(z)\, \ln(1-z)\right]\, \right.\nonumber\\
&&\left.\hspace{-2cm}+  \mathlarger{\int}_x^{z_{\mbox{\tiny max}}} dz \, \hat{P}_{qq}^{(0)}(z)\,\ln_{\mbox{\tiny LxR}}+ \ln_{\mbox{\tiny LxR, max}}\, \mathlarger{\int}_{z_{\mbox{\tiny max}}}^1 dz \, \hat{P}_{qq}^{(0)}(z)\,
\right\} \frac{x}{z}\, q\left(\frac{x}{z},Q^2\right). \nonumber \\
\quad 
\label{eq:convo_max}
\end{eqnarray}   
where the PDFs are included under the integral.
Our definition of the maximum value for the argument of the running coupling follows from the realization of duality in the resonance region. The value $z_{max}$ is reached at
\begin{eqnarray}
R^{\mbox{\tiny exp/th}}(z_{\mbox{\tiny max}}, Q^2)
&=&\frac{
 \mathlarger{\int}_{x_{\mbox {\tiny min}}}^{x_{\mbox {\tiny max}}} dx\,
F_2^{\mbox {\tiny exp}} (x, Q^2)
}
{ \mathlarger{\int}_{x_{\mbox{\tiny min}}}^{x_{\mbox{\tiny max}}} dx\,
F_2^{NS, \mbox{\tiny Resum}} (x, z_{\mbox{\tiny max}}, Q^2)
}
= \frac{I^{\mbox{\tiny exp}}}{I^{ \mbox{\tiny Resum}}}= 1\quad.
\label{eq:ratio_max}
\end{eqnarray}
 The results are depicted by the red hexagons on Fig.~\ref{fig:ratio_lxr}. 
%

\section{The Running Coupling Constant from LxR}
\label{sec:alpha}

\begin{figure}[h]
\includegraphics[scale= .7]{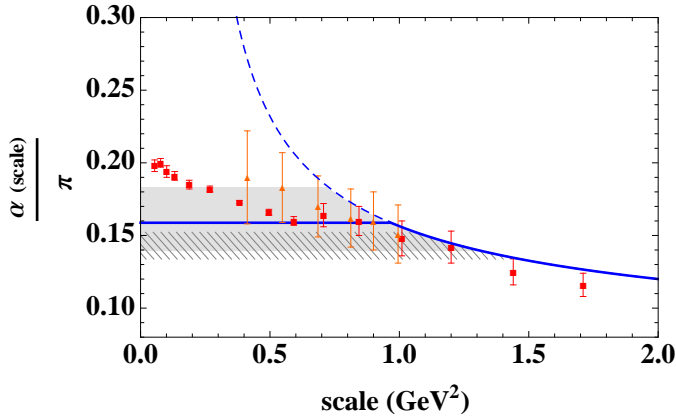}
\caption{Extraction of $\alpha_s$. The blue dashed curve represents the exact NLO solution for the running coupling constant in $\overline{\mbox{MS}}$ scheme. The solid blue curve represents the coupling constant obtained from our analysis using inclusive electron scattering data at large $x$.  Owing to large $x$ resummation, at  lower values of the scale, $\alpha_s=\alpha_{s, \mbox{\tiny NLO}}\left(\mbox{min} \right)$ is frozen as explained in the text. The grey area represents the region where the freezing occurs for JLab data, while the hatched area corresponds the freezing region determined from SLAC data. This  error band represents  the theoretical uncertainty in our analysis. We also plot results extrapolated from the recent analysis of A.~Deur: the red squares correspond to $\alpha_s$ extracted from Hall B CLAS EG1b, with statistical uncertainties; the  orange triangles corresponds to Hall A E94010 / CLAS EG1a
data, the uncertainty here contains both statistics and systematics.
\label{extract}}
\end{figure}

The direct consequence of the previous Section is that duality is realized, within our assumptions, by allowing $\alpha_s$  to run from a minimal  scale only. From that minimal scale downward to the real photon limit (scale=0GeV$^2$), the coupling constant does not run, it is frozen. This feature is illustrated on Fig.~\ref{extract}. We show the behavior of $\alpha_{s, \mbox{\tiny NLO}}$(scale) in the $\overline{\mbox{MS}}$ scheme and for the same value of $\Lambda$ used throughout this paper. 
The theoretical errorband correspond to the extreme values of
\begin{equation}
\alpha_{s, \mbox{\tiny NLO}}\left(Q_i^2\frac{(1-z_{max, i})}{z_{max, i }}\right) \qquad \mbox{for}\qquad i=1,\ldots 10 \qquad,
\end{equation}
$i$ corresponds to the data points.
Of course, we expect the transition from nonperturbative to perturbative to occur at one unique scale. The discrepancy between the 10 values we have obtained has to be understood as the resulting error propagation. The grey area represents the approximate frozen value of the coupling constant,
\begin{equation}
0.13\leq \frac{\alpha_{s, \mbox{\tiny NLO}} (\mbox{scale}\to 0 \mbox{GeV}^2)}{\pi} \leq 0.18 \quad.
\end{equation}

%
In the figure we also report  values from the extraction using polarized $eP$ scattering data in Ref.~[\refcite{Deur:2005cf,Deur:2008rf,Brodsky:2010ur,alexandre}]. These values represent the  first extraction of an effective coupling in the IR region that was obtained by analyzing the data  
relevant for the study of the GDH sum rule. To extract the coupling constant,  the $\overline{\mbox{MS}}$ expression of the Bjorken sum rule up to the 5th order in alpha (calculated in the $\overline{\mbox{MS}}$ scheme) was used. In order to compare with our extraction using the $F_2^p$ observable, the finite value for $\alpha_s(0)$  found in [\refcite{Deur:2005cf,Deur:2008rf,Brodsky:2010ur}] was rescaled in [\refcite{alexandre}] assuming the validity of the commensurate scale relations [\refcite{Brodsky:2010ur}] in the entire range of the scale entering the analysis. 
The agreement with our analysis which is totally independent, is impressive.

\section{Conclusions}

We report an interesting observation that  the values of the coupling from different measurements/observables 
namely the GDH sum rule, and our large-$x$-DIS/resonance region based extractions, 
are in very good agreement with the values obtained from the extension of the commensurate scale relations [\refcite{Grunberg:1980ja,Brodsky:1994eh}] to the IR region suggested in Ref.~[\refcite{Deur:2008rf,alexandre}].  
The issue of the extension of the scheme/observable dependence to low values of the scale is what makes our new extraction interesting and open to further studies.
While our conclusion ensues from a perturbative analysis, in the near future, we will explore the role of non-perturbative 
effective couplings as well.

The importance of finite couplings has been highlighted many times in the Literature.
Our analysis allows  to extract from a fit~[\refcite{us}] the nonperturbative parameters often present in the proposed functional forms for the running of $\alpha_s$, {\it e.g.}, in Refs.~[\refcite{Cornwall:1981zr,Fischer:2003rp,Shirkov:1997wi}].

\section*{Acknowledgments}

The support of Gruppo III of INFN, Laboratori Nazionali di Frascati, where part of the manuscript was completed is wholeheartedly acknowledged. This work was funded by the Belgian Fund F.R.S.-FNRS via the contract of ChargŽe de recherches (A.C.), and by U.S. D.O.E. grant  DE-FG02-01ER4120 (S.L.).


\begin{thebibliography}{0}    




\bibitem{Bloom:1970xb} E. D. Bloom and F. J. Gilman, {\it Phys.Rev.Lett.} {\bf 25}, 1140 (1970). 
\bibitem{Bloom:1971ye} E. D. Bloom and F. J. Gilman, {\it Phys.Rev. D} {\bf 4}, 2901 (1971).
\bibitem{Liuti:2011rw} S. Liuti, {\it Int.J.Mod.Phys.Conf.Ser.} {\bf 04}, 190 (2011).
\bibitem{Bianchi:2003hi} N. Bianchi, A. Fantoni, and S. Liuti, {\it Phys.Rev. D} {\bf 69}, 014505 (2004).
\bibitem{Courtoy:2013qca} A.~Courtoy and S.~Liuti, (2013)  arXiv:1302.4439 [hep-ph].
\bibitem{Rujula:1976tz} A. De Rujula, H. Georgi, and H. D. Politzer, {\it Annals Phys.} {\bf 103}, 315 (1977).
\bibitem{Melnitchouk:2005zr} W. Melnitchouk, R. Ent, and C. Keppel, {\it Phys.Rept.} {\bf 406}, 127 (2005).
\bibitem{Liang:2004tj} Y. Liang et al. (Jefferson Lab Hall C E94-110 Collaboration) (2004), nucl-ex/0410027.
\bibitem{Liuti:2001qk} S. Liuti, R. Ent, C. Keppel, and I. Niculescu, {\it Phys.Rev.Lett.} {\bf 89}, 162001 (2002). 
\bibitem{Monaghan:2012et} P. Monaghan, A. Accardi, M. Christy, C. Keppel, W. Melnitchouk, et al. (2012), arXiv:1209.4542.
\bibitem{Whitlow:1991uw} L. Whitlow, E. Riordan, S. Dasu, S. Rock, and A. Bodek, {\it Phys.Lett. B} {\bf 282}, 475 (1992).
\bibitem{Martin:2009iq} A. Martin, W. Stirling, R. Thorne, and G. Watt, {\it Eur.Phys.J. C} {\bf 63}, 189 (2009).
\bibitem{Georgi:1976ve} H. Georgi and H. D. Politzer, {\it Phys.Rev. D} {\bf 14}, 1829 (1976).
\bibitem{Niculescu:2000tk} I. Niculescu, C. Armstrong, J. Arrington, K. Assamagan, O. Baker, {\it et al.}, {\it Phys.Rev.Lett.} {\bf 85}, 1186 (2000). 
\bibitem{Poggio:1975af} E. Poggio, H. R. Quinn, and S. Weinberg, {\it Phys.Rev. D} {\bf 13}, 1958 (1976).
\bibitem{Amati:1980ch} D. Amati, A. Bassetto, M. Ciafaloni, G. Marchesini, and G. Veneziano, {\it Nucl.Phys. B} {\bf 173}, 429 (1980).
\bibitem{Roberts:1999gb} R. Roberts, {\it Eur.Phys.J. C} {\bf 10}, 697 (1999).
\bibitem{Niculescu:1999mw} I. Niculescu, C. Keppel, S. Liuti, and G. Niculescu, {\it Phys.Rev. D} {\bf 60}, 094001 (1999).

\bibitem{Deur:2005cf} A. Deur, V. Burkert, J.-P. Chen, and W. Korsch,  {\it Phys.Lett. B} {\bf 650}, 244 (2007).
\bibitem{Deur:2008rf} A. Deur, V. Burkert, J. Chen, and W. Korsch, {\it Phys.Lett. B} {\bf 665}, 349 (2008).
\bibitem{Brodsky:2010ur} S. J. Brodsky, G. F. de Teramond, and A. Deur,  {\it Phys.Rev. D} {\bf 81}, 096010 (2010).
\bibitem{alexandre} A. Deur, Private Communication (2013).
\bibitem{Grunberg:1980ja} G. Grunberg, {\it Phys.Lett. B}{\bf 95}, 70 (1980),  G. Grunberg, {\it Phys.Rev. D} {\bf 29}, 2315 (1984).
\bibitem{Brodsky:1994eh} S. J. Brodsky and H. J. Lu,  {\it Phys.Rev. D} {\bf 51}, 3652 (1995).
\bibitem{us} A. Courtoy, S. Liuti, and V. Vento, In progress (2013).
\bibitem{Cornwall:1981zr} J. M. Cornwall, {\it Phys.Rev. D} {\bf 26}, 1453 (1982).

\bibitem{Fischer:2003rp} C. S. Fischer and R. Alkofer,  {\it Phys.Rev. D} {\bf 67}, 094020 (2003), hep-ph/0301094.
\bibitem{Shirkov:1997wi} D. Shirkov and I. Solovtsov,  {\it Phys.Rev.Lett.} {\bf 79}, 1209 (1997), hep-ph/9704333.
\end{thebibliography}
\end{document}